%% file: main.tex
\documentclass[conference]{IEEEtran}

\usepackage{amsmath,amsfonts}
\usepackage{algorithmic}
\usepackage{graphicx}
\usepackage{adjustbox} 
\def\BibTeX{{\rm B\kern-.05em{\sc i\kern-.025em b}\kern-.08em
    T\kern-.1667em\lower.7ex\hbox{E}\kern-.125emX}}
\usepackage{url}
\usepackage{hyperref}
\usepackage{multirow}
\usepackage{makecell}
\usepackage{booktabs}

\usepackage[group-separator={,},group-minimum-digits=4, group-digits=integer]{siunitx}

\usepackage[skins]{tcolorbox}
\newcommand{\answer}[1]{\begin{tcolorbox}
[enhanced,colback=white,boxrule=0.5pt,boxsep=4pt,left=0pt,right=0pt,top=0pt,bottom=0pt,after={\vspace{-0.1cm}}]
{#1}\end{tcolorbox}}

\newboolean{showcomments}
\setboolean{showcomments}{true}

\ifthenelse{\boolean{showcomments}}
  {\newcommand{\nb}[3]{
  {\color{#2}\small\fbox{\bfseries\sffamily\scriptsize#1}}
  {\color{#2}\sffamily\small$\triangleright~$\textit{\small #3}$~\triangleleft$\GenericWarning{}{LaTeX Warning: #1: #3}}
  }
  \newcommand{\todo}[1]{{\color{red}{TODO: #1}}\GenericWarning{}{LaTeX Warning: TODO: #1}}
  }
  {\newcommand{\nb}[3]{}
  \newcommand{\todo}[1]{}
  }

\usepackage{listings}
\usepackage{color}
\usepackage[font=footnotesize]{caption}
\definecolor{lightgray}{rgb}{.9,.9,.9}
\definecolor{darkgray}{rgb}{.4,.4,.4}
\definecolor{purple}{rgb}{0.65, 0.12, 0.82}

\lstdefinelanguage{JavaScript}{
  keywords={typeof, new, true, false, catch, function, return, null, catch, switch, var, let, const, if, in, while, do, else, case, break},
  keywordstyle=\color{blue}\bfseries,
  ndkeywords={class, export, boolean, throw, implements, import, this},
  ndkeywordstyle=\color{darkgray}\bfseries,
  identifierstyle=\color{black},
  sensitive=false,
  comment=[l]{//},
  morecomment=[s]{/*}{*/},
  commentstyle=\color{purple}\ttfamily,
  stringstyle=\color{red}\ttfamily,
  morestring=[b]',
  morestring=[b]"
}

\begin{document}

\title{On the Bug-proneness of Structures Inspired by Functional Programming in JavaScript Projects\vspace{-6pt}}

\author{\IEEEauthorblockN{Fernando Alves\IEEEauthorrefmark{1}, Delano Oliveira\IEEEauthorrefmark{1}, Fernanda Madeiral\IEEEauthorrefmark{2}, and Fernando Castor\IEEEauthorrefmark{3}}
\IEEEauthorblockA{\IEEEauthorrefmark{1}Federal University of Pernambuco, Recife, Brazil, \{fhaa, dho\}@cin.ufpe.br}
\IEEEauthorblockA{\IEEEauthorrefmark{2}KTH Royal Institute of Technology, Stockholm, Sweden, fer.madeiral@gmail.com}
\IEEEauthorblockA{\IEEEauthorrefmark{3}Utrecht University, Utrecht, Netherlands, f.j.castordelimafilho@uu.nl}
\vspace{-16pt}
}

\maketitle

\begin{abstract}
Language constructs inspired by functional programming have made their way into most mainstream programming languages. Many researchers and developers consider that these constructs lead to programs that are more concise, reusable, and easier to understand. Notwithstanding, few studies investigate the prevalence of these structures and the implications of using them in mainstream programming languages. This paper quantifies the prevalence of four concepts typically associated with functional programming in JavaScript: recursion, immutability, lazy evaluation, and functions as values. We divide the latter into two groups, higher-order functions and callbacks \& promises. We focus on JavaScript programs due to the availability of some of these concepts in the language since its inception, its inspiration from functional programming languages, and its popularity. We mine \num{91} GitHub repositories (more than \num{22} million LOC) written mostly in JavaScript (over \num{50}\% of the code), measuring the usage of these concepts from both static and temporal perspectives. We also measure the likelihood of bug-fixing commits removing uses of these concepts (which would hint at bug-proneness) and their association with the presence of code comments (which would hint at code that is hard to understand). We find that these concepts are in widespread use (\num{478605} occurrences, \num{1} for every \num{46.65} lines of code, \num{43.59}\% of LOC). In addition, the usage of higher-order functions, immutability, and lazy evaluation-related structures has been growing throughout the years for the analyzed projects, while the usage of recursion and callbacks \& promises has decreased. We also find statistical evidence that removing these structures, with the exception of the ones associated to immutability, is less common in bug-fixing commits than in other commits. In addition, their presence is not correlated with comment size. Our findings suggest that functional programming concepts are important for developers using a multi-paradigm language such as JavaScript, and their usage does not make programs harder to understand.
\end{abstract}

\section{Introduction}

Functional programming is a programming paradigm where programs are built by defining, applying, and composing functions~\cite{scott2016programming}. Many researchers~\cite{Hudak:1989:CEA, Hughes:1989:WFP, Backus:1978:CPL} consider that functional concepts lead to more concise, reusable, and easier to understand programs. Elm, Scheme, and Haskell are examples of functional programming languages. Multi-paradigm languages, such as JavaScript and Python, also include constructs popularized by functional languages. JavaScript, for example, takes inspiration from functional languages like Lisp and Scheme~\cite{saternos2014client}. By mixing different paradigms, these languages allow one to solve problems in an easier or more efficient way than one could do with a single paradigm.

The extent to which developers use functional programming concepts to develop in multi-paradigm programming languages is unknown. Perhaps developers are not even aware they are using these concepts when programming with these languages. Moreover, the use of functional programming concepts in a language that is not purely functional can be confusing. Consider callbacks, for example, which are functions passed as arguments to other functions and executed after them. They induce a non-linear control flow and can be deferred to execute asynchronously, declared anonymously, and may be nested to arbitrary levels, which can be challenging to understand and maintain~\cite{Gallaba2015}. Studying how developers use constructs inspired by functional programming in a mainstream multi-paradigm programming language can show us how these constructs are used in practice and provide subsidies for researchers and tool builders to propose improvements. 

In this work, we investigate the use of concepts inspired by the functional programming paradigm in JavaScript, a mainstream, imperative, multi-paradigm language. In particular, our main goal is to examine the bug-proneness of changes that employ these structures. To achieve this goal, we mine \num{91} open-source repositories in JavaScript and collect data about the usage of \num{22} functional programming structures related to the concepts of recursion, immutability, lazy evaluation, and functions as values. To the best of our knowledge, this is the first study to investigate the usage of JavaScript constructs inspired by functional programming concepts.

We find out that \num{43.59}\% of the $\sim$22M analyzed lines of code are related to functional programming structures, and it is possible to find one use of such structures for every \num{46.65} lines of code. Furthermore, the usage of higher-order functions, immutability, and lazy evaluation-related structures has been growing throughout the years for the analyzed projects, while the usage of recursion and callbacks \& promises has decreased. In particular, the usage of immutability-related structures has more than tripled throughout their evolution. Moreover, we find out that the concepts tend to be removed less often in bug-fixing commits than in non-bug-fixing commits, excepting immutability-related structures, and that there is no correlation between comment size and source code including them. These findings highlight the importance of functional programming structures to developers even in an inherently imperative language and suggest that the usage of these structures is less error-prone than not using them and does not make the source code difficult to understand.

\noindent
\textbf{Artifacts availability.} The data analyzed in this work and the tools used to obtain and process it are publicly available~\cite{anonymous_2022_6425005}.

\section{Functional programming concepts}\label{sec:fps}

Functional programming is a programming paradigm where programs are built by defining, applying, and composing functions~\cite{scott2016programming}. Unlike imperative programming, which has roots in the Turing machine, functional programming has its roots in the Lambda Calculus~\cite{Church:1936:UPE}. In a functional programming language, functions are first-class values, i.e., they can be assigned to variables, passed as arguments, and returned from functions. Furthermore, functional literals can be defined. Functional languages typically include several features not generally available in imperative languages, such as lazy evaluation, partial function application, and absence of side effects. If a functional language treats every computation as the evaluation of a mathematical expression, i.e., it does not allow for state changes and mutable data, it is considered pure. Although initially focused mainly on academic research, functional programming has seen a surge in popularity in the industry in the last 15 years. Some widely popular languages, such as Python, Swift, and JavaScript, include constructs popularized by functional languages since their first versions. In addition, many mainstream imperative languages, such as C\#, Java, and C++, have introduced constructs and libraries to support a programming style heavily inspired by functional languages.

This section presents which concepts related to functional programming we take into account in our study. It is important to note that these concepts are considered elements of the functional programming paradigm by multiple authors~\cite{scott2016programming, watt2004programming, kereki2020mastering}. However, they do not represent everything possible in the languages of this paradigm. Some concepts can be broken down into several (e.g., currying, pure functions, monads), and we do not always seek to analyze them. In \autoref{sec:mining}, we detail how these concepts are identified in the analyzed systems.

\vspace{5pt}
\noindent\textit{Recursion}: A function is recursive if it calls itself directly or indirectly~\cite{hinsen2009promises}. Although this is not specific to functional programming languages, recursion is an important concept because, in the absence of side effects, it provides the only general means of performing repetition~\cite{scott2016programming}.

\vspace{5pt}
\noindent\textit{Immutability}: Immutable data is a standard feature of functional programming, and it is used to prevent changes (mutation) in data structures. In other words, it means the value of an expression depends only on the referencing environment in which it is evaluated and not on the time at which the evaluation occurs. If an expression yields a specific value at one point in time, it is guaranteed to yield the same value at any point in time (referential transparency)~\cite{scott2016programming}. Immutability, when applied to objects, for example, creates objects that cannot be modified after they have been created~\cite{brady2021immutability}. Notice that immutability is an essential concept to purely functional languages since, for those languages, there are no side effects.

\vspace{5pt}
\noindent\textit{Lazy evaluation}: This concept is related to the idea that an expression is only evaluated when its value is necessary, e.g., because it is used in an I/O operation. Lazy evaluation is a counterpoint to eager evaluation, where an operator is applied as soon as its operands are known~\cite{watt2004programming}. With lazy evaluation, it is possible, for example, to avoid unnecessary calculations and create infinite lists.

\vspace{5pt}
\noindent\textit{Functions as values}: In functional languages, functions are values, which means that they can be passed as parameters, returned from subroutines, or assigned to variables~\cite{scott2016programming}. In addition, it is possible to define function literals, also known as anonymous functions or lambdas. A function that takes another function as an argument or returns a function as its result is called a higher-order function~\cite{scott2016programming}. This concept is derived from mathematics and can be intuitively considered a function of functions~\cite{xu2020mining}. An example of a higher-order function is \texttt{map}, which takes a list \textit{l} and a function \textit{f} as its arguments and applies \textit{f} to each element of \textit{l}.

\section{Study design}

\subsection{Research questions}\label{sec:rqs}

To have a clearer understanding of the use of concepts inspired by functional programming in JavaScript programs, we formulated research questions that would indicate whether the phenomenon in question is a commonplace situation or not. In addition, we are also looking to answer questions that could relate the use of these structures to associated bug-fix rates and comments. Therefore, we focus on the following questions:

\vspace{5pt}
\newcommand\rqone{How often are functional programming concepts used in real software?}
\noindent\textbf{RQ1}. \rqone

\vspace{2pt}
\newcommand\rqtwo{How has the use of functional programming concepts evolved over the years?}
\noindent\textbf{RQ2}. \rqtwo

\vspace{2pt}
\newcommand\rqthree{Are uses of functional programming concepts removed more often in bug-fixing commits?}
\noindent\textbf{RQ3}. \rqthree

\vspace{2pt}
\newcommand\rqfour{Is code that employs functional programming concepts associated to longer comments?}
\noindent\textbf{RQ4}. \rqfour

\vspace{5pt}

RQ1 aims to gauge the pervasiveness of functional programming concepts in real-world JavaScript software. RQ2 provides a temporal perspective, measuring the evolution in using these concepts in the studied projects. For RQ3, we are interested in assessing bug-proneness. Although it is difficult to identify the causes, we can use the code that changes when bugs are fixed as a proxy~\cite{Sliwerski:2005:WCI}. Therefore, the rationale for RQ3 is that if bug-fixing commits are more likely to remove instances of functional programming concepts than non-bug-fixing commits, this may indicate that these concepts are bug-prone. Finally, the rationale behind RQ4 is that, since developers write comments to help them better understand the functioning and the intent of code snippets, we expect code with longer comments to be more troublesome to understand than code with shorter comments, as reported in previous work~\cite{Aman:2015:LCN, Steidl2013, Buse2008}. With this question, we seek to understand whether there is an association between comment length and functional programming concepts.

\subsection{Project selection}

To answer our research questions, we first select the repositories to be analyzed. We aim to select a representative sample of mature repositories written mainly in JavaScript. We also want our sample to follow good GitHub data-mining practices. More specifically, as mentioned by Hinsen~\cite{hinsen2009promises}, the selected projects should be active, i.e., with at least one commit in the last six months, at least \num{1000} commits, and \num{1000} issues overall, not personal or archived projects, and created more than six months ago. We use GitHub because of its popularity among developers, documented ways of accessing its content through APIs, and the possibility of accessing open-source software (OSS) from diverse domains.

Unfortunately, GitHub does not offer ways of directly obtaining a list of repositories using the criteria mentioned earlier. It provides channels to search for repositories through its search API\footnote{\url{https://docs.github.com/en/rest/reference/search}} and using the advanced search\footnote{\url{https://github.com/search/advanced}} but not without having to develop a software system to group the returned data. So, after testing some methods~\cite{krishna2018connection} and tools (e.g., Reaper~\cite{munaiah2017curating}), we decided to use a tool called SeArt\footnote{\url{https://seart-ghs.si.usi.ch/}}~\cite{dabic2021sampling}, because it allows us to get a list of repositories with the mentioned criteria.

We use the following settings to search repositories using SeArt: \texttt{Language} (JavaScript), \texttt{Number of commits} (\num{1000}, at minimum), \texttt{Number of contributors} (\num{2}, at minimum, to avoid personal repositories), \texttt{Number of issues} (\num{1000}, at minimum), \texttt{Created at} (before 2021-01-01, six months before collection date), \texttt{Last commit at} (after 2021-01-01, six months after collection date) and \texttt{Exclude forks} (yes). This search returns \num{357} repositories. We then remove the archived ones, not found (git cloning returned a \texttt{not found} error), and the ones whose GitHub topics were related to documentation or guides. After this filtering step, we keep \num{338} repositories. 

With these results, we run a tool called \texttt{cloc}\footnote{\url{https://github.com/kentcdodds/cloc}} to obtain the Count of Lines of Code (CLOC) of each repository because we use this information in some research questions and also to prioritize repository processing. We ignore some folders (\texttt{node\_modules}, \texttt{coverage}, \texttt{build}, \texttt{bin}, \texttt{stories}, \texttt{dist} and \texttt{3rdParty}) trying to avoid code that is usually related to the build process, third-party libraries, or auto-generated code. Despite searching only for JavaScript projects, we also consider code written in TypeScript. According to its website\footnote{\url{https://www.typescriptlang.org/}}, TypeScript is just \textit{``JavaScript with syntax for types''}. In other words, we accept all extensions related to JS and TS as options for \texttt{cloc}. Executing this tool resulted in a count of \num{35396336} lines of code. Finally, we order the repository list by CLOC and select the \num{100} projects with the most LOC. Since we could not parse nine of them, our final list has \num{91} projects amounting to \num{22326070} LOC. \autoref{tab:projects} summarizes some statistics of the selected projects. It shows that our sample comprises mature repositories with extensive histories in the number of commits and diverse in several aspects such as popularity (i.e., number of stars) and number of contributors.

\input{tables/table_projects}

\subsection{Mining functional programming concepts' usages in JavaScript programs}\label{sec:mining}

When mining for the usage of functional programming concepts, we search for specific blocks of code that can represent those concepts described in \autoref{sec:fps}. To process them, we developed a tool that can build an AST (Abstract Syntax Tree) from the files of each repository and recognize elements from the source code of these systems as functional programming structures. Our tool was developed by extending the TypeScript Compiler API because it allows us to infer types in a way that other parsers we have tested (e.g., Esprima\footnote{\url{https://esprima.org/}}) were not able to, adding more reliability to the results.

We also verify the precision of the results produced by our tool in two ways. First, we draw a random sample of \num{280} code snippets and manually check whether the classification the tool performs is correct. To reduce the impact of disproportionately large projects, we first randomly select a project and then randomly pick one code snippet including a potential functional programming structure from that project. We repeat this procedure \num{280} times. In this step, we did not find any misclassifications. Second, to ensure that every structure is represented, since some of them appear only rarely, e.g., the \texttt{flatMap} function, we randomly select five instances of each structure we have considered, across all the projects, totalling \num{105} code snippets. Also, we did not find any misclassification in this step. Table~\ref{tab:prevalence} presents the complete list of structures.

In the remainder of this section, we explain the JavaScript structures that we have selected to represent the functional programming concepts described in \autoref{sec:fps}.

\vspace{2pt}
\noindent\textbf{Recursion.}
We looked for function declarations whose names are used as call expressions once or more inside their bodies. We consider only direct recursion, i.e, we do not account for cases where a function \textit{f} calls a function \textit{g} which calls \textit{f}. Previous work has shown that indirectly recursive calls are uncommon~\cite{Carter:2018:EEQ}. Furthermore, this type of analysis would considerably increase the processing time, since it would require the identification of cycles in the program call graph.

\vspace{2pt}
\noindent\textbf{Immutability.}
When parsing for immutability, we look for structures representing the idea of copying a data structure (object or array) shallowly or preventing it from being changed. We do not, for example, look for deep clones or libraries related to immutability (e.g., Ramda, Underscore.js). Therefore, we consider four scenarios for immutability.

The first case we consider is the use of the \texttt{Object.freeze} method. It prevents an object (and its prototype) from being changed. When analyzing the repositories, we used the TypeScript type checker to infer when a call expression has a left-hand side expression with an object constructor, and its name is \texttt{freeze}, to reduce the probability of false positives.

Next, we consider the use of spread syntax for immutability because it is used to (shallowly) copy or destructure arrays and objects without modifying them. To process these structures, we search for array literal expressions that represent spread elements and spread assignments. \autoref{code:spread-syntax} shows an example of an object (\texttt{person}) being shallowly copied into another object (\texttt{anotherPerson}). This new object has its \texttt{age} property changed to 51, but that does not affect the first object.

\begin{lstlisting}[caption={Example of spread syntax.},label={code:spread-syntax}]
const person = { age: 50 };
const anotherPerson = { ...person, age: 51 };
console.log(person.age); // 50
console.log(anotherPerson.age); // 51
\end{lstlisting}
\vspace{-0.6em}

In addition to the spread syntax for shallow copies, we also look at two other structures with the same purpose, \texttt{Object.assign} (with an empty object passed in the first parameter) and \texttt{Array.slice} (with no arguments taken). In the first situation, we look for a call expression with precisely two arguments where the first one is an empty object and the second one is an object. We also take uses of \texttt{Array.slice} into account because, when no arguments are taken, the \texttt{slice} function returns a copy of an array in its entirety, in a similar way to those previously mentioned. Parsing this structure requires only searching for call expressions from arrays whose names are \texttt{slice} with zero arguments.

\vspace{2pt}
\noindent\textbf{Lazy evaluation.}
Although JavaScript does not support lazy evaluation inherently, it includes mechanisms that can delay the evaluation of an expression (or execution of a statement) until it is necessary. This work examines two such mechanisms, named generator functions and thunks. 

Generator functions are not directly inspired by functional programming~\cite{kereki2020mastering}. Notwithstanding, we consider them a way of achieving lazy evaluation because instead of immediately processing an expression when invoked, these functions return a particular type of iterator called generator. This iterator only has its value consumed when the generator's \texttt{next} method is called, executing the function until it finds the \texttt{yield} keyword. With generators, it is possible, for example, to create infinite lists, a typical example of the uses of lazy evaluation~\cite{Hughes:1989:WFP}. An asterisk token can identify generator functions or methods in their declarations.

In \autoref{code:generator}, there is a generator function that returns numeric values every time its \texttt{next} method is called. It is important to note that, thanks to lazy evaluation, it is possible to use an infinite data structure (\texttt{Infinity}) without running out of memory.

\begin{lstlisting}[caption={Example of a generator function.},label={code:generator}]
function* range() {
    let count = 0;
    for (let i = 0; i < Infinity; i++) {
        count++;
        yield i;
    }
    return count;
}
const iterator = range();
console.log(iterator.next().value); // 0
console.log(iterator.next().value); // 1
\end{lstlisting}
\vspace{-0.6em}

A thunk is a nullary function literal, i.e., an arrow function that has no parameters. Thunks encapsulate computations that are only executed when they are actually invoked. As a consequence, they can also be used to delay evaluation~\cite{kereki2020mastering}. In \autoref{code:thunk}, the expression ``2 + 2'' is only evaluated when the thunk \texttt{four} is called.

\begin{lstlisting}[caption={Example of a thunk.},label={code:thunk}]
const four = () => 2 + 2;
\end{lstlisting}
\vspace{-0.6em}

\vspace{2pt}
\noindent\textbf{Functions as values.}
Due to the emphasis of this work on JavaScript, we divide functions as values into two groups: \textbf{higher-order functions (HOFs)} and \textbf{callbacks \& promises}.

When parsing for HOFs, we look for two specific scenarios. The first scenario occurs when a function takes another function as an argument and uses it to traverse a list applying it to each component of the list~\cite{watt2004programming}. To identify this scenario, we look for function names that refer to native functions from \texttt{Array.prototype} (\texttt{every}, \texttt{filter}, \texttt{find}, \texttt{findIndex}, \texttt{flat}, \texttt{flatMap}, \texttt{forEach}, \texttt{map}, \texttt{reduce}, \texttt{reduceRight} and \texttt{some}), and receive functions as arguments to traverse a list. We search for type-inferred arguments that are arrow functions, function expressions, or type-inferred functions. We ignore cases where TypeScript is unable to infer the type of the argument.

The second scenario consists of non-native functions (created by a developer) returning another function as their result. We disregard non-native function declarations that take functions as parameters due to a limitation of the Typescript compiler API that does not infer functions types in that manner. This does not include callbacks, which we address later in this section. In addition, there are many ways to call functions in JavaScript but we decided to only parse \texttt{property access} expressions, i.e., of the form \texttt{o.f()}. We are also not considering \texttt{Array.prototype} overriding. We discuss these limitations in more detail in \autoref{sec:threats}.

Callbacks are functions passed as an argument to another (parent) function and they are typically used in asynchronous calls such as timeouts and \texttt{XMLHttpRequests} (\texttt{XHRs}). Callbacks are executed after the parent function has completed its execution~\cite{fard2013jsnose}, that is, ``as a handler to be called in response to some future event.''~\cite{scott2016programming}. In this work, we only look for functions that use callbacks, that is, functions whose parameter names are called once or more inside their bodies. More specifically, we consider scenarios where declared functions are passed as arguments or where the argument is an anonymous function.

A promise is an object that represents the eventual completion (or failure) of an asynchronous operation and its resulting value. We consider it an excellent example of native use of functions as values because when promises are created, it is possible to pass at least two arguments (two functions) that will be called when a promise succeeds (\texttt{resolve}) or fails (\texttt{rejects}). Promises became popular in JavaScript as an approach to discipline the use of callbacks. To parse them, we look only for declarations of promises, so we type-check \texttt{new} expressions creating objects of type \texttt{Promise}. \autoref{code:promise} shows a promise that fulfills with the value \texttt{1} when it resolves, printing it in the console.

\begin{lstlisting}[caption={Example of a promise.},label={code:promise}]
new Promise((resolve) => {
	resolve(1)
}).then(console.log) // 1
\end{lstlisting}
\vspace{-1.5em}
\subsection{Methods}\label{sec:methods}

To answer RQ1, we first have to identify all uses of functional programming concepts from the main revision of each repository. To do this, we visit every AST node searching for specific node kinds that can include uses of these concepts, as explained in the previous section. This process is done in conjunction with the data extraction of RQ2, except that, for RQ1, we only consider the revision marked with GitHub's main revision SHA.

To get data to answer RQ2, we go through each commit for each repository, analyzing one snapshot (due to computational cost) per month when available. This process starts on the main revision date and ends with the first available commits made in the repositories. For each set of functional programming concepts we identify, we save this data to use during the analysis step.

To address RQ3, we need to obtain commits and classify them as bug-fixing commits or non-bug-fixing commits. To do so, we rely on labels provided by developers to categorize issues and pull requests on GitHub repositories. We use the REST and GraphQL GitHub APIs to fetch issues and pull requests based on queried labels related to bugs and then obtain the commits that closed these issues and pull requests for every analyzed project. 

First, we define a list of terms that are related to bugs, which is composed of \texttt{bug}, \texttt{error}, \texttt{defect}, \texttt{failure}, \texttt{fault}, and \texttt{exception}. These terms are used in a query to fetch labels related to bugs in a given repository. In this matching process, we ignore labels that contain one bug-related term together with the term \texttt{unconfirmed} or \texttt{not}.

After that, we start the process for each repository included in our study. We first get all labels from a given repository and classify them as bug-related labels or not according to the terms described above. Then, we fetch issues and pull requests from the repository because when we were developing our extraction tool, there was no way of directly downloading commits based on the labels of the associated issues and pull requests. Thus, for each bug-related label in the repository, we download up to \num{1000} closed issues or pull requests since the last repository commit date using a GraphQL query. We chose not to recursively download all issues and pull requests because it would take much longer to get all the data available in each repository, which would also add considerable processing time to the data analysis. We only take into account the issues that were closed by a commit or by a pull request through the GraphQL GitHub closing event (\texttt{CLOSED\_EVENT}) so that we can associate them to their fixing commits. Finally, from the selected issues and pull requests, we fetch the last commit associated with them and consider them as bug-fixing commits. To download non-bug-fixing commits, we follow the same steps, but the query to fetch the issues and pull requests is made so as not to bring data that contains the bug-related labels. We consider that the commits found in these issues and pull requests are not from bug fixes.

Finally, our tool analyzes the repository versions from the bug-fixing commits and the non-bug-fixing commits to identify and count how many functional programming structures were removed. When bug-fixing commits removed functional programming structures, we consider that the removal is part of a bug fix. Note that this analysis is performed on the complete snapshots of the identified commits and their parents, not the difference between them. We consider the entire snapshot because we leverage the available information to improve the precision of the detection of some functional programming structures. Having only the parts of the code that were modified limits TypeScript's ability to infer types.

To answer RQ4, we also visit every AST node on each repository's main git revision, looking for comments. To parse them, we use the TypeScript Compiler API to collect information about the positioning of the comments, based on ranges from the full text of each source file and node positioning (\texttt{full start} for leading comments and \texttt{end} for trailing comments), their types (\texttt{leading} or \texttt{trailing}), whether they are adjacent to functional programming structures or not, and whether they have JSDoc tags. In addition, we also remove repeated comments, as the same comment can appear in more than one AST node and persist them in CSV files.

\section{Study results}

In this section, we present the results of the study. Each one of the four RQs is addressed in its respective section. 

\subsection{\rqone} %Functional programming concepts' usage frequency (RQ1)}

In our corpus of \num{91} open-source JavaScript projects, we identified \num{478605} occurrences of functional programming structures. Considering that the analyzed projects have \num{22326070} lines of code, that means that there is one use of functional programming concepts for every \num{46.65} lines of code, on average.
Table~\ref{tab:prevalence} shows a breakdown of these occurrences in terms of these structures (column ``Occurrences (\#)'') and the percentage of the overall LOC of the analyzed projects related to functional programming (column ``\% LOC'').

If we examine the number of lines of code of the functional programming structures, \num{43.59}\% of all the lines of code in the projects are related to functional programming. It means that almost one out of every two lines in these projects is related to functional programming. This number may sound inflated, but it makes sense when we consider that, for example, for a higher-order function, we include all the lines of the function in this count. The rationale for this conservative approach is to account for the code affected by functional programming in its entirety. For example, for the code snippet in Listing~\ref{code:promise}, we would count three lines of code. It may lead to some lines of code being counted more than once for different structures, e.g., a callback may use higher-order functions and invoke \texttt{Object.freeze}. This is the reason why, if we add up all the percentages in the column ``\% LOC'' of Table~\ref{tab:prevalence}, the result will be greater than the aforementioned \num{43.59}\%. Different approaches could have been employed, but then identifying which parts of the code pertain to functional programming concepts and which ones do not become fuzzy. 
For example, in an anonymous function used as a callback, should we ignore its body when counting the lines of code? The answer is not clear. We mitigate this problem by presenting both the number of LOC and the number of occurrences of each structure.

In total, the most pervasive concept is lazy evaluation, with \num{299520} occurrences. The least pervasive one is recursion, with only \num{7879} occurrences. When we consider the structures related to the concepts, the most and least frequent ones are, respectively, \texttt{thunks}, with \num{298797} instances, and the higher-order function \texttt{reduceRight}, with only \num{25} occurrences. Furthermore, we did not find occurrences of \texttt{flat} functions. When we consider the number of lines of code, callbacks \& promises comprise \num{4168527} LOC, i.e., \num{18.67}\% of all the LOC in the analyzed projects. Immutability, the functional programming concept with the least lines of code, comprises only \num{172462} LOC, i.e., \num{0.77}\% of all the LOC.

Table~\ref{tab:prevalence} also presents the proportions of all the occurrences of functional programming concepts represented by each structure. Furthermore, the sum of the occurrences of thunks and callbacks accounts for \num{71.66}\% of all the occurrences of functional programming structures in the analyzed projects.

\begin{figure}[bt]
    \centering
    \includegraphics[scale=0.25]{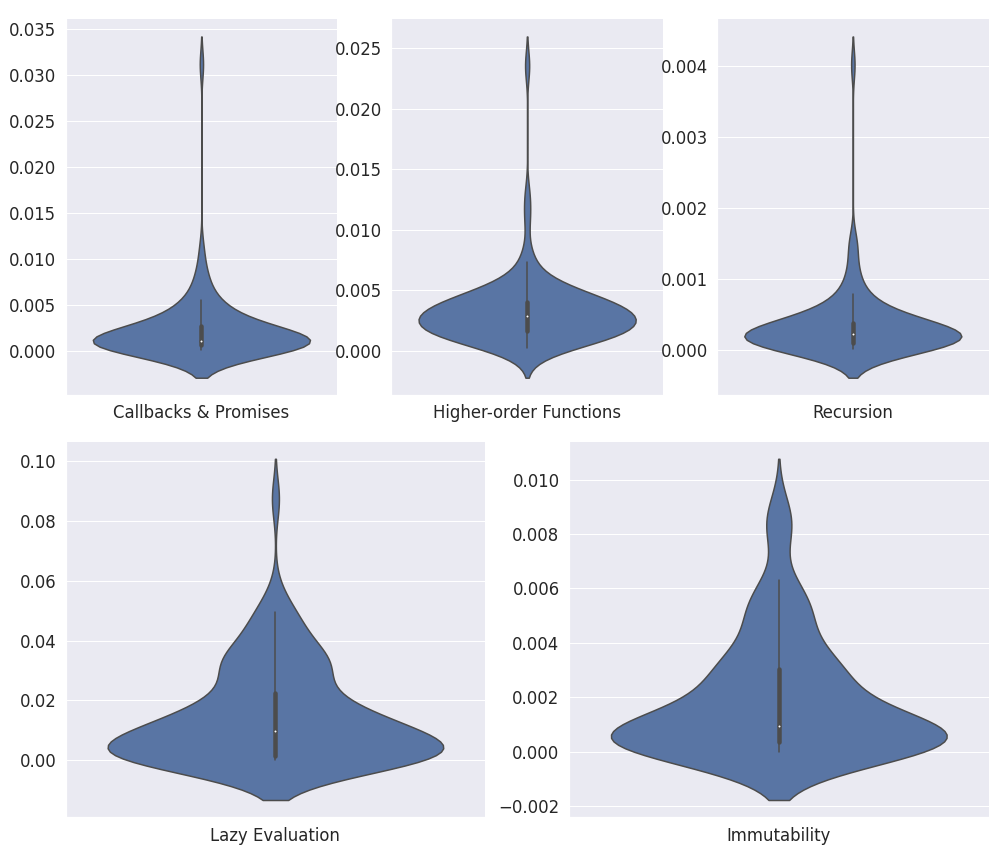}
    \vspace{-2em}
    \caption{Distribution of use of functional programming concepts in repositories}
    \label{fig:rq1_violin_repos}
    \vspace{-1.5em}
\end{figure}

\autoref{fig:rq1_violin_repos} shows the distributions of the usages of functional programming concepts in the repositories. The LOC of the projects normalizes the distributions. Note that the scale is different for each violin plot. The usage of callbacks \& promises, higher-order functions, and recursion is overall consistent for the projects, which is observed by the density of projects in the median of the distributions. However, this consistency does not exist with immutability-related structures and the lazy-evaluation-related ones.

\input{tables/table_rq1}

\answer{
{%\small
\textbf{Key takeaways for RQ1}. %\rqone \\
We found out that functional programming concepts are used very often in the analyzed projects, on average, once for every \num{46.65} lines of code. In addition, the code related to these structures comprises \num{43.59}\% of all the LOC in these projects. The most popular of these structures are \texttt{thunks}. They represent \num{62.43}\% of all functional programming structures and occur six times more often than the second most popular one, callbacks.}
}

\subsection{\rqtwo }\label{sec:rq2}

We also investigate how the use of functional programming concepts changes throughout the evolution of the repositories. We adopt the following approach. First, for each monthly snapshot (see Section~\ref{sec:methods}) of each project, we count the number of lines of the functional programming concepts, normalized by the overall number of LOC of the project. Second, we calculate the percentage change for each pair of consecutive monthly snapshots. If any of the two snapshots has a zero value, we discard it and use the next one with a non-zero value. Finally, we compute the geometric mean of all the percentage changes for each repository and structure. The geometric mean is useful to summarize changes in percentages over time. For example, if the geometric mean of the percentage changes for one of the functional programming concepts, for one project, is exactly \num{1.0}, this means that the use of that concept did not change throughout the evolution of that project. In case it is \num{1.01}, this means that, on average, the use of that concept grew \num{1}\% per subsequent snapshot. Since the values are normalized by the number of LOC, this is a real growth in the use of functional programming.

\begin{figure}[bt]
    \centering
    \includegraphics[trim={4cm 0cm 4cm 4cm},clip, scale=0.20]{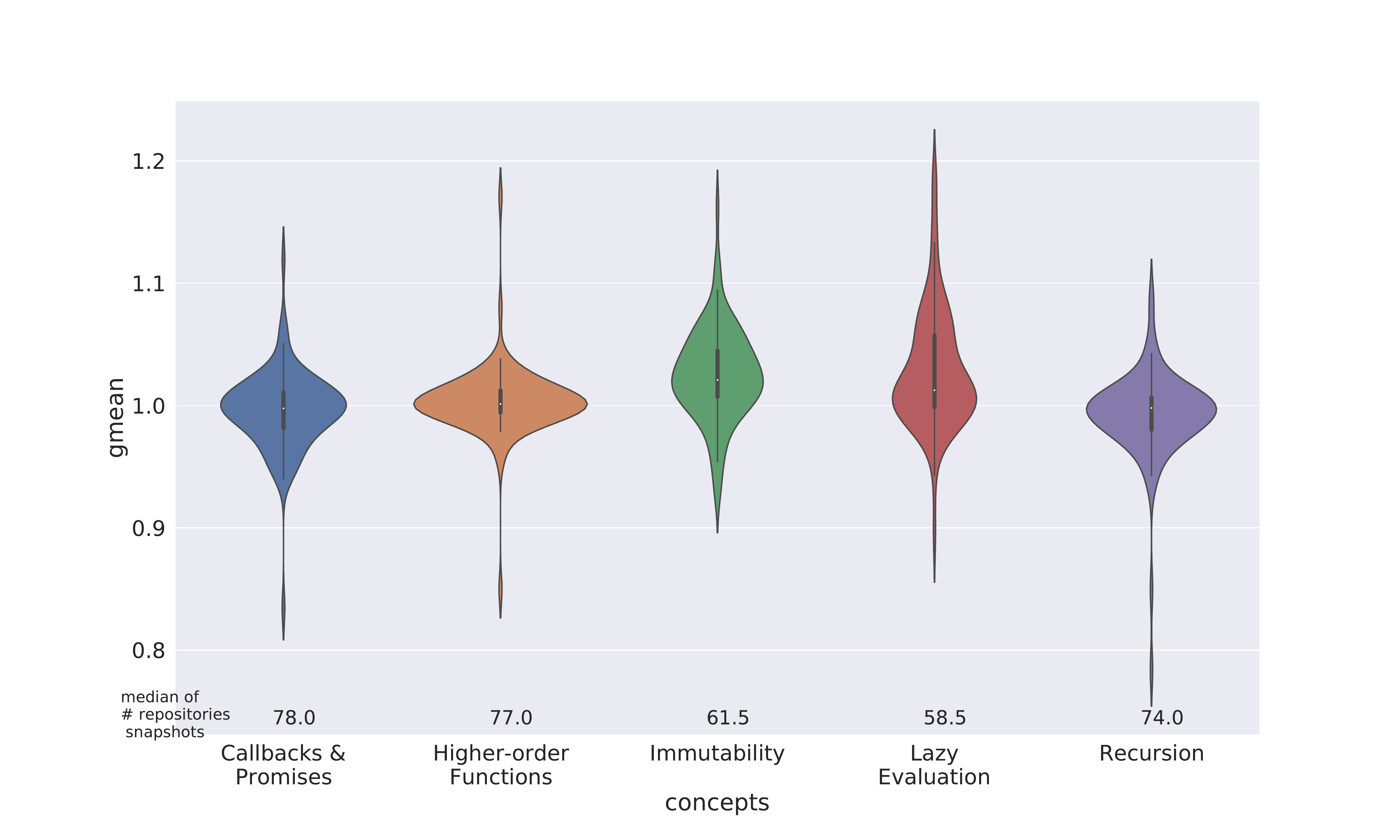}
    \vspace{-0.9em}
    \caption{Distribution of evolution of functional programming concepts in repositories}
    \vspace{-1.5em}
    \label{fig:rq2_violinplot_gmean}
\end{figure}

The violin plot in \autoref{fig:rq2_violinplot_gmean} presents the result of the processing. Each violin shows the distribution of the geometric means of the analyzed projects for each functional programming concept. The white dot in each violin indicates the median geometric mean. The numbers at the bottom of the plot show the median number of snapshots based on the geometric means calculated. The plot shows that all the medians sit close to \num{1.0}. Table~\ref{tab:gmeans} presents the order statistics for the data in the plot. It is possible to see that \num{50}\% of the projects exhibited a monthly growth of \num{2.0825}\% in the use of immutability structures. Since we are considering \num{61.5} snapshots (as shown at the bottom of the plot), this represents an overall growth of \num{255}\% ($1.0208258^{61.5}$) in the use of immutability-related structures, on average. The usage of two other functional programming concepts also grew in the analyzed projects: lazy evaluation (\num{107.21}\%) and higher-order functions (\num{9.74}\%). For the two remaining concepts, there was a reduction. Overall usage of recursion fell by \num{13.67}\% and callbacks \& promises by \num{16.65}\%.

\answer{
{%\small
\textbf{Key takeaways for RQ2}. %\rqtwo \\
The use of functional programming structures has been growing throughout the years for the analyzed projects. However, this growth is uneven and inconsistent for all functional programming concepts. Usage of immutability-related structures has grown by more than \num{255}\% whereas usage of constructs for lazy evaluation has doubled. On the other hand, the usage of recursion and callbacks \& promises has decreased by 13.67\% and 16.65\%, respectively. 
}}

\subsection{\rqthree}

To analyze the error-proneness of functional programming concepts, we start from the null hypothesis that there is no relationship between bug-fixing commits and the removal of functional programming structures. More specifically, we formulate five null hypotheses, one for each functional programming concept. We perform the chi-square test considering two dimensions: bug-fixing vs. non-bug-fixing commits and with vs. without removal of functional programming structures. We apply the Bonferroni adjustment to the alpha since we test five hypotheses. Thus, we use an alpha of \num{0.01}.

Table~\ref{tab:bugs} presents the obtained p-values for the chi-square test. In all comparisons, the null hypotheses are rejected, excepting immutability. In other words, there is a relationship between the removal of functional programming structures and bug-fixing commits but not for the latter. For all the tests that rejected the null hypothesis, the p-values are orders of magnitude lower than \num{0.01}.

We then calculate the odds ratio to quantify the odds of functional programming structures removed in a bug-fix commit. Table~\ref{tab:bugs} shows the results for the functional programming concepts. For example, the odds ratio for recursion is \num{0.622}. It indicates that functional programming structures are \num{37.77}\% \textbf{less} likely to be removed in bug-fixing commits than in non-bug-fixing commits. This same phenomenon can be observed for all the cases that rejected the null hypothesis, although less intense. Structures related to lazy evaluation, higher-order functions, and callbacks \& promises are \num{4.16}\%, \num{6.89}\% and \num{17.38}\% less likely to be removed in bug-fixing commits, respectively. These results suggest that using functional programming structures in JavaScript programs is less bug-prone than not using them, except for immutability.

\input{tables/table_gmeans}

\answer{
{%\small
\textbf{Key takeaways for RQ3}. Functional programming structures tend to be removed less often in bug-fixing commits than in non-bug-fixing commits. It can be observed for all the functional programming concepts, excepting immutability. The difference is more prominent for recursion and callbacks \& promises than for other functional programming concepts. They are \num{37.77}\% and \num{17.38}\% less likely to be removed in bug-fixing commits, respectively.}}

\subsection{\rqfour}

Code comments aim to make code easier for developers by explaining how it works or the rationale behind its leading design and implementation decisions. Some authors~\cite{Beck:1999:EPE, Hunt:1999:PP, Fowler:2019:RID} argue that comments indicate problems with the code associated with, especially if they appear within methods. Fowler~\cite{Fowler:2019:RID} states the following about code comments: 

\begin{quote}
\textit{It is surprising how often you look at thickly commented code and notice that the comments are there because the code is bad.}
\end{quote}

\noindent We investigate whether comments associated with code that leverages functional programming concepts tend to be longer than comments for code that does not include these structures. We expect longer comments associated with code that is harder to understand~\cite{Aman:2015:LCN, Steidl2013, Buse2008}. Our sample has \num{1644133} comments, \num{17772} of them are adjacent to the functional programming structures and \num{1626361} are not. Of these comments, \num{131342} are trailing and \num{1512791} are leading. \num{311365} are using JSDoc tags. The most commented concept is higher-order functions with \num{13842} comments. The one with the least is immutability.

We use the point-biserial correlation coefficient to check the correlation between comment size (a continuous variable) and the presence of functional programming structures (a dichotomous variable) in the associated code snippet. We do not take JSDoc tag comments into account as they have a specific structure used to generate documentation for coarse-grained entities, e.g., entire methods or classes. \autoref{tab:comments} shows the results of the point-biserial correlation for each of the functional programming concepts. For all the concepts the correlations are negligible, and the p-values suggest that it is not possible to infer a relationship between comment size and the use of functional programming concepts. Considering the combination of all structures, without distinguishing between the concepts, we obtain a p-value of \num{0.510} and a correlation of \num{-57.107e-05}. This correlation is negligible, and the p-value indicates no statistical significance. This result suggests that code including functional programming concepts is not more challenging to understand than code that does not. 

\answer{
{%\small 
\textbf{Key takeaways for RQ4}. We found no correlation between comment size and code with functional programming concepts. When examining the different functional programming concepts separately it is not possible to ascertain whether there is a relationship or not. }
}

\section{Discussion}

The JavaScript language includes support for higher-order functions and enables every function to be treated as a value since its first version in 1995. More recent versions of the ECMAScript specification, e.g., ES 6, in 2015, extend this support with constructs such as arrow functions and the \texttt{const} declaration. This study shows that these constructs have widespread adoption in the analyzed projects. At the same time, besides the use of callbacks~\cite{Gallaba2015, Gallaba:2017:RAJ, Saburoy:2017:ESC}, we are not aware of any other study in the literature that studies how ideas from functional programming influence software development in JavaScript. Researchers have an opportunity to fill in this knowledge gap by investigating in-depth topics such as how developers use immutability structures, how to support these developers in building (mostly) purely functional programs in JavaScript, and how to refactor existing systems to leverage these structures.

Not only are functional programming structures in widespread use, but their use has also been growing even when we normalize that growth based on the number of lines of code in each project. As pointed out in Section~\ref{sec:rq2}, on average, the frequency of occurrence of structures related to lazy evaluation, mainly thunks, has grown by more than \num{100}\% throughout the life of these projects. Although there was a reduction in recursion and callbacks \& promises, these reductions were comparatively small. In the former case, the reduction may be explained by the growth in the use of high-order functions, since many of these functions (\texttt{map}, \texttt{filter}, \texttt{reduce}) perform operations that are typically implemented recursively. In the latter case, part of the reduction can be explained by the use of alternative constructs, such as \texttt{async-await}, introduced in ECMAScript 8 (2017), which provide a more disciplined way of using promises. Similarly, the growth in the use of thunks can be partially explained by the publication of the ECMAScript 6 specification, which introduced these structures. Notwithstanding, it does not explain why \num{62}\% of all the occurrences of functional programming structures in projects pertain to this case. Investigating these changes in tendencies in more depth is left for future work.

\input{tables/table_rq3_correlation}

\input{tables/table_rq4_correlation}

An important point raised during the execution of this research was whether we should consider \texttt{const} declarations or not. A \texttt{const} declaration can be seen as connected to immutability because it declares a block-scoped, read-only variable. So, the value of a \texttt{const} variable will not be reassigned or redeclared within the same scope because its reference is immutable. Despite this, the Mozilla Developer Network Web Docs\footnote{\url{https://developer.mozilla.org/}} consider that \texttt{const} is not an immutability structure because such as variable may be assigned  an object or array, which is mutable. If we hypothetically consider this element to be related to the functional programming concept of immutability in our analyses, it would significantly impact the results. For example, it becomes the most pervasive structure, with \num{900858} instances. About one in every \num{25} lines of code in the analyzed projects would consist of a \texttt{const} declaration. Also, almost two out of every three occurrences of functional programming structures would be uses of \texttt{const} declarations. The sum of the occurrences of \texttt{const} declarations, thunks, and callbacks would account for \num{90}\% of all the occurrences of functional programming structures. Furthermore, the usage of immutability structures would have grown, on average, by approximately \num{400}\% throughout the evolution of the analyzed projects.

Since the early days of functional programming, it has been touted to as a way to write code that is clearer and easier to understand. In his Turing Award Lecture~\cite{Backus:1978:CPL} in 1978, John Backus remarked about a functional program that \textit{``its structure is helpful in understanding it without mentally executing it''}. A decade later, Hughes~\cite{Hughes:1989:WFP} argued that \textit{``[functional programming] allows improved modularization.''}. According to him, mechanisms such as lazy evaluation and high order functions make it possible to write simpler programs by decomposing them into small, easy to write and read pieces. Hudak~\cite{Hudak:1989:CEA} emphasized the importance of immutability, arguing that \textit{``although the notion of referential transparency may seem like a simple idea, the clean equational reasoning that it allows is very powerful, not only for reasoning formally about programs but also informally in writing and debugging programs.''} These authors and their ideas have been very influential in the Programming Languages and Software Engineering communities. Even though they focus on statically-typed, purely functional languages, this paper shows that there is potential benefit in leveraging these ideas, even if only partially, in the context of a dynamically-typed, imperative programming language. Our study provides evidence that the use of structures inspired by functional programming does not make code harder to understand while being potentially less bug-prone. Considering that functional programming structures have seen little empirical evaluation, this is an important step that can motivate the community to conduct more studies.

Finally, we tried to understand if there is any correlation between comment length and the usage of functional programming-inspired constructs in the neighboring code, but this is not the only aspect to be considered. The presence of Self-Admitted Technical Debt (SATD) comments has been analyzed by some works \cite{Bavota2016, Maldonado2017}. Analyzing whether there is any relationship between SATD comments and the use of functional programming is another dimension for future work. In addition, an in-depth analysis can also be performed on the removal of these comments.

\section{Threats to validity}\label{sec:threats}

\vspace{5pt}
\noindent\textit{Construct validity}. Some structures cannot be directly related to their concept. For example, we considered \textit{spread} as a structure that promotes immutability because developers use it to create copies of objects and arrays when they do not want to change the original ones. However, someone can use it to create copies for another intention that is not immutability.

Some ways of mining structures were not considered because they were not representative enough. For example, there are many ways to call functions in JavaScript, but, after sampling four GitHub popular projects (React, Angular, ESLint, and Hoodie), we discovered that only \num{0.07}\% of function call expressions were not \texttt{property access expressions}, so we decided to ignore other access expressions, e.g., \textit{element access expression}, when parsing. We also searched for \texttt{Prototype} overriding of higher-order functions, e.g., redefinitions of function \texttt{map}, and found no cases in our random sample. It means that it is possible to identify native higher-order JavaScript functions by their name and target object type.

Furthermore, as in most implementations of source code analysis, our parser has some limitations. For example, we did not find any case of \texttt{flat} functions. An in-depth analysis would show the reasons to it. Moreover, in a literal array where each element is in one line, and there is a comment in each line, the parser only returns the comment in the last line of the literal array. 
Also, due to the nature of the structures we mine and the high dynamism of the JavaScript language, the precision of the mining process may be negatively affected. We gauge the precision of the tool we have built by manually checking the results produced by it, as reported in Section~\ref{sec:mining}. We did not find any misclassifications.

\vspace{5pt}
\noindent\textit{Internal validity}.
To mine the usage of functional programming structures, we analyzed the entire projects' versions to find additions and removals instead of the actual changes between pairs of versions. Analyzing the changes would increase the engineering effort for implementing the parser and make it impossible to mine some structures. However, by analyzing the entire versions of the projects, we do not have the mapping of the actual structures from one version to another, which is a threat to our study. For instance, if a given existing structure in a version of the project was deleted in a subsequent one, but in that subsequent version, a different structure, but of the same kind, was added, our parser will count it as the same structure.

Moreover, for RQ3, we hypothesize that if functional programming structures are deleted in bug-fixing commits, their usage might be bug-prone. This hypothesis assumes that all changes in a bug-fixing commit are about the bug fix. However, research projects have shown that bug-fixing commits contain other changes that are not related to bug fixes~\cite{Herbold2021}.

Finally, for RQ4, we compared the size of comments in source code related to functional programming structures with the size of comments in other source code. Our analysis is performed at the AST level, and a comment might belong to more than one AST node since comments are at the line level. It would bias our results because the same comment would be counted for several nodes. To solve that, each comment found in multiple AST nodes was considered only for the first node. However, the existing threat to our study's validity is that a comment might belong to AST nodes related to functional programming structures and other code, and we cannot know for sure what the comment is about. In such a case, we keep the comment for one AST node of each type of code (related and non-related to functional programming structures).

\vspace{5pt}
\noindent\textit{External validity}. Even though we searched for projects from various domains, it is still possible that our sample is not representative enough among the many repositories available on GitHub. Furthermore, it is impossible to relate our findings with enterprise software development, mainly because we analyzed only open-source software. Moreover, the functional programming structures we choose may not be representative when generalizing functional languages. As we decided to focus on some scenarios, we know that we do not evaluate several other functional programming structures, so we can not say how representative these other structures would be.

\section{Related work}

We organize related work in terms of two main lines: studies on the usage of functional programming (Section~\ref{sec:rw:functional}) and mining studies targeting JavaScript projects (Section~\ref{sec:rw:js}). 

\subsection{Usage of functional programming structures}\label{sec:rw:functional}

We found in the literature studies investigating the usage of specific functional programming structures. For example, Gallaba et al.~\cite{Gallaba2015} investigated the usage of callback in a corpus of \num{138} JavaScript programs. They found out that, on average, every \num{10}th function definition takes a callback as an argument. Xu et al.~\cite{xu2020mining} analyzed the use of high-order functions in Scala programs. They collected \num{8285} higher-order functions from \num{35} Scala projects and found out that \num{6.84}\% of functions are defined as higher-order functions on average. Figueroa et al.~\cite{Figueroa2021} analyzed the use of monads as a dependency in \num{85135} packages in the Haskell language. They found that \num{32}\% of the packages depend on the packages that implement monads. Mazinanian et al.~\cite{Mazinanian:2017:UUL}
analyzed the usage of lambda expressions in \num{241} open-source Java projects. The authors found out that the ratio of lambdas introduced per added line of code increased by \num{54}\% between 2015 and 2016 and also discovered that developers adopt lambdas for reasons such as making code more concise and avoiding duplication. We investigated more than one functional programming concept in our work, differently from those studies that focused on one specific concept. 
%Furthermore, the usage of parameter-less lambda expressions (thunks) also grew significantly for the period we considered. In the work of Mazinanian et al., they considered the growth between 2015 and 2016, while lambda expressions had been introduced in Java in March 2014. In our case, 

Following along different lines, Lubin et al.~\cite{Lubin:2021:HST} studied how programmers write code in several statically-typed functional programming languages, including Haskell, Elm, F\#, and others. The authors conducted a grounded theory analysis of \num{30} programming sessions, combined with \num{15} semi-structured interviews, and produced a theory of how programmers write code in these languages. They then validated some of the elements of that theory in a controlled experiment and found out, for example, that programmers in these languages tend to code in a cycle of writing a bit of code and running the compiler, even when it is clear that compilation will fail. Furthermore, pattern matching tended to incur a reduced workload compared to combinators. 

Kamps et al.~\cite{Kamps:2019:AQE} studied structural degradation in Haskell programs by monitoring three static metrics related to size, cohesion, and coupling. The authors leveraged the Gini coefficient to measure structural inequality. They found out that post-release defects correlate significantly with the degree of inequality between the size of the modules in three mature Haskell systems.

\subsection{Mining JavaScript projects}\label{sec:rw:js}

We found some studies that mined JavaScript repositories. Hanam et al.~\cite{Hanam2016} mined 105K commits from 134 server-side JavaScript projects aiming to discover bug patterns. In a paper by Campos et al.~\cite{Campos2019}, they mined code snippets in JavaScript on Stack Overflow to analyze them using ESLinter, a JavaScript linter. Furthermore, they investigated the use of those code snippets in GitHub projects. Saboury et al.~\cite{Saboury2017} investigated code smells in \num{537} releases of five popular JavaScript applications aiming to understand how they impact the fault-proneness of applications. They detected \num{12} types of code smells (e.g., nested callbacks and variable re-assign) and found out that, on average, files without code smells have hazard rates \num{65}\% lower than files with code smells. 

Richards et. al~\cite{Richards:2011:EMD} conducted a large-scale study of the use of the \texttt{eval} function in JavaScript-based web applications. They recorded the behavior of \num{337}MB of strings given as arguments to \num{550358} calls to \texttt{eval} exercised in over \num{10000} websites and observed that, at the time, between \num{50}\% and \num{82}\% of the most popular websites used \texttt{eval}. They also confirmed, in that context, that \texttt{eval} usage is pervasive and not necessarily something problematic.

\section{Conclusion}

This paper presents a study on how JavaScript programs employ concepts inspired by functional programming. We have analyzed \num{91} JavaScript projects amounting to more than \num{22} million lines of code. Our investigation has revealed that the projects employ functional programming structures intensively: they occur, on average, once for every \num{46.65} lines of code, and more than \num{54}\% of all the lines of code in these projects are related in some way to these structures. Furthermore, their adoption is growing intensively. Immutability- and lazy evaluation-related structures have exhibited growths of \num{255}\% and \num{107}\% along with the evolution of the analyzed projects. We also found that functional programming structures, excepting the immutability-related ones, tend to be removed less often in bug-fixing commits than in non-bug-fixing commits. For callbacks \& promises and recursion, occurrences of these concepts are \num{17.38}\% and \num{37.77}\% less likely to be removed in bug-fixing commits, respectively. Finally, we did not find a relationship between comment size and the use of functional structures.

\bibliographystyle{IEEEtran}
\bibliography{references}

\end{document}

%% file: tables/table_projects.tex
\begin{table}[]
\caption{Selected projects.}\label{tab:projects}
\centering
{\footnotesize
\begin{tabular}{@{} @{\hspace{0.2\tabcolsep}} l@{\hspace{0.1\tabcolsep}} rrrrr @{}}
    \toprule
    {} & \multicolumn{1}{c}{Min} & \multicolumn{1}{c}{25\%} & \multicolumn{1}{c}{50\%} & \multicolumn{1}{c}{75\%} & \multicolumn{1}{c}{Max} \\
    \midrule
    LOC               & \num{101952}  & \num{116077}  & \num{173106}  & \num{284617}  & \num{1857932} \\
    \# Stargazers     & \num{17}      & \num{380}     & \num{2982}    & \num{17840.5} & \num{171203} \\
    \# Contributors   & \num{13}      & \num{76}      & \num{171}     & \num{346}     & \num{1504} \\
    \# Commits        & \num{1765}    & \num{5924}    & \num{9768}    & \num{17849}   & \num{77667}\\
    \# Issues         & \num{1007}    & \num{1878}    & \num{2879}    & \num{6939}    & \num{21538}\\
    \# Pull requests  & \num{1002}    & \num{1775.5}  & \num{2971}    & \num{5089}    & \num{36763}\\
    \bottomrule
\end{tabular}
}
\vspace{-2em}
\end{table}

%% file: tables/table_rq1.tex
\begin{table}[t]
	\caption{Prevalence of functional programming structures.}\label{tab:prevalence}
	\centering
	\footnotesize
	\begin{tabular}{@{}l l r r@{}}
		\toprule
		Concept & Structure & \% LOC & Occurrences (\num{478605}) \\
		\midrule
		Higher-order
		        & every             & 0.0133\%   & 0.1272\%     (\num{609}) \\
		functions
		        & filter            & 0.0753\%   & 1.0971\%     (\num{5251}) \\
		{}      & find              & 0.0181\%   & 0.4461\%     (\num{2135}) \\
		{}      & findIndex         & 0.0021\%   & 0.0501\%     (\num{240}) \\
		{}      & flat              & 0\%        & 0\%          (\num{0}) \\
		{}      & flatMap           & 0.0010\%   & 0.0079\%     (\num{38}) \\
		{}      & forEach           & 0.6475\%   & 2.9024\%     (\num{13891}) \\
		{}      & map               & 0.3179\%   & 2.1454\%     (\num{10268}) \\
		{}      & reduce            & 0.0879\%   & 0.4643\%     (\num{2222}) \\
		{}      & reduceRight       & 0.0005\%   & 0.0052\%     (\num{25}) \\
		{}      & some              & 0.0160\%   & 0.2513\%     (\num{1203}) \\
		{}      & Non-native        & 8.8892\%   & 7.4418\%     (\num{35617}) \\
		{} & \textit{Total}         & 10.0688\%  & 14.9388\%    (\num{71499}) \\
		\midrule
		Immutability
		        & Array.slice       & 0.0034\%   & 0.1529\% (\num{732}) \\
		{}      & Object.assign     & 0.0067\%   & 0.2171\% (\num{1039}) \\
		{}      & Object.freeze     & 0.0526\%   & 0.3661\% (\num{1752}) \\
		{}      & Spread Assignment & 0.6263\%  & 5.4366\% (\num{26020}) \\
		{}      & Spread Element    & 0.0916\%  & 1.0403\% (\num{4979}) \\
		{}      & \textit{Total}    & \num{0.77}\% & 7.213\% (\num{34522}) \\
		\midrule
		Callbacks \&
		        & Callback          & 17.8525\%     & 9.2330\%  (\num{44190}) \\
		Promises
		        & Promise           & 0.8767\%      & 4.3867\%  (\num{20995}) \\
		{} & \textit{Total}         & \num{18.67}\% & 13.6197\% (\num{65185}) \\
		\midrule
		Lazy
		        & Generator         & 0.0435\%      & 0.1510\% (\num{723}) \\
		evaluation
		        & Thunk             & 15.3588\%     & 62.4308\% (\num{298797}) \\
		{}      & \textit{Total}    & 15.4023\%     & 62.5818\% (\num{299520}) \\
		\midrule
		Recursion
		        & {}                & 1.2233\%    & 1.6462\% (\num{7879}) \\
		\bottomrule
	\end{tabular}
	\vspace{-1em}
\end{table}

%% file: tables/table_gmeans.tex
\newcommand{\relativeSpace}{0.85}
\begin{table}[t]
	\caption{Order statistics for the geometric means summarizing the evolution in the use of functional programming structures in the analyzed projects. A value of 1.0 means no change.}\label{tab:gmeans}
	\centering
	{\footnotesize
		\begin{tabular}{@{} @{\hspace{0.2\tabcolsep}} l@{\hspace{\relativeSpace\tabcolsep}} r@{\hspace{\relativeSpace\tabcolsep}} r@{\hspace{\relativeSpace\tabcolsep}} r@{\hspace{\relativeSpace\tabcolsep}} r@{\hspace{\relativeSpace\tabcolsep}}
		r @{}}
			\toprule
			Concept     & \multicolumn{1}{c}{Min} & \multicolumn{1}{c}{25\%} & \multicolumn{1}{c}{50\%} & \multicolumn{1}{c}{75\%} & \multicolumn{1}{c}{Max} \\
			\midrule
			Callbacks \& promises  & 0.834839  & 0.981676  & 0.997667  & 1.010751  & 1.119621    \\
			Higher-order functions & 0.849989  & 0.994403  & 1.001208  & 1.012228  & 1.170697    \\
			Immutability           & 0.926663   & 1.007126  & 1.020825  & 1.044866  & 1.161784  \\
			Lazy evaluation        & 0.895239   & 0.998629  & 1.012532  & 1.057329  & 1.185875  \\
			Recursion              & 0.783807   & 0.980235  & 0.998015  & 1.006601  & 1.089953  \\
			\bottomrule
		\end{tabular}
	}
	\vspace{-1em}
\end{table}

%% file: tables/table_rq3_correlation.tex
\begin{table}[t]
\caption{The p-values and odds ratios for the relationship between bug-fixing commits and the removal of functional programming structures. For all the cases, the removal of functional programming structures is \textbf{less} likely to occur in bug fixing commits than in non-bug fixing commits.}\label{tab:bugs}
\centering
\footnotesize
\begin{tabular}{@{} llr @{}}
    \toprule
    Concept                 & Chi-square test (p-value)   & Odds ratio       \\
    \midrule
    Recursion                    & \num{4.580e-93}          & \num{0.622}   \\
    Lazy evaluation              & \num{2.591e-07}         & \num{0.958}  \\
    Higher-order functions       & \num{5.049e-14}         & \num{0.931}  \\
    Callbacks \& Promises        & \num{1.196e-52}         & \num{0.826}  \\
    Immutability                 & 0.015276            &                           \\
    \bottomrule
\end{tabular}
%\vspace{-1em}
\end{table}

%% file: tables/table_rq4_correlation.tex
\begin{table}[t]
\caption{Correlations between usage of functional programming structures and code comment size. The p-values only indicate statistical significance for functional programming in general. For all the cases, correlation was negligible.}\label{tab:comments}
\centering
\footnotesize
\begin{tabular}{@{} llr @{}}
    \toprule
    Concept                & Point-biserial coefficient             & p-value                      \\
    \midrule
    Recursion                   & \num{-12.064e-05}                 & \num{0.889}               \\
    Lazy evaluation             & \num{-13.906e-05}                 & \num{0.872}              \\
    Higher-order functions      & \num{-50.456e-05}                   & \num{0.560}               \\
    Callbacks \& Promises       & \num{-23.197e-05}                  & \num{0.789}               \\
    Immutability                & \num{-6.026e-05}                  & \num{0.944}              \\
    All                         & \num{-57.107e-05}                  & \num{0.510}              \\
    \bottomrule
\end{tabular}
\vspace{-1.5em}
\end{table}